\documentclass[
    jap,
    reprint,
    showkeys,
    amsmath,
    amssymb
]{revtex4-1}

\usepackage[utf8]{inputenc}

\usepackage{graphicx}
\usepackage{dcolumn}
\usepackage{bm}
\usepackage{siunitx}

\usepackage{cleveref}

\renewcommand{\epsilon}{\varepsilon}
\renewcommand{\phi}{\varphi}

\let\origcite\cite%
\def\cite#1{\unskip~\origcite{#1}}

\begin{document}

\title{%
    Steady-state and transient behavior in dynamic atomic force microscopy%
}

\author{Tino Wagner}
\email{E-mail: tiwagner@ethz.ch}

\affiliation{%
    Nanotechnology Group,
    ETH Zürich, Säumerstrasse 4, 8803 Rüschlikon, Switzerland
}

\date{\today}

\begin{abstract}
    We discuss the influence of external forces on the motion of the tip in
    dynamic atomic force microscopy (AFM).
    First, a compact solution for the steady-state problem is derived employing
    a Fourier approach.
    Founding on this solution, we present an analytical framework to describe
    the transient behavior of the tip after perturbations of tip--sample forces
    and the excitation signal.
    The static and transient solutions are then combined to obtain the baseband
    response of the tip, i.e., the deflection signal demodulated with respect to
    the excitation.
    The baseband response generalizes the amplitude and phase response of the
    tip, and we use it to find explicit formulas describing the amplitude and
    phase modulation following the influence of external forces on the tip.
    Finally, we apply our results to obtain an accurate dynamic model of the
    amplitude controller and phase-locked loop (PLL) driving the cantilever in a
    frequency modulated AFM setup. A special emphasis is put on discussing the
    tip response in environments of high damping, such as ambient or liquid.
\end{abstract}

\maketitle

\section{Introduction}

Atomic force microscopy is an extremely versatile microscopy technique in
surface science, which utilizes the force between a tip moving above the surface
to obtain a highly resolved topographic image. Since its invention in
1986\cite{Binnig:1986vr}, there have been numerous improvements to the
technique. First and foremost, there has been a transition to dynamic operation
modes\cite{Garcia:2002kc}, in which the tip, mounted on a force sensor such as a
cantilever, is forced to oscillate near or at the resonance frequency. The
tip--sample force changes the resonant behavior of the force sensor, allowing
one to detect changes of the tip--sample interaction from changes of the
detected oscillation amplitude\cite{Martin:1987ht} (amplitude modulated, AM-AFM)
or resonance frequency\cite{Albrecht:1991bu} (FM-AFM) of the cantilever.

In addition to topography measurements, complementary information can be
obtained from the tip--surface interaction. For example, the power dissipated by
the tip can be measured and related to viscoelastic damping and frictional
forces\cite{Cleveland:1998cn,Anczykowski:1999da}. The tip--sample force also
contains magnetic\cite{Martin:1987fw}, electrostatic\cite{Martin:1988ec}, and
near-field optical\cite{Nowak:2016gs} information about the sample. Individual
force contributions can be modulated by an external stimulus, such as an
external voltage bias or pulsed-light illumination, to facilitate separation
from other contributions in the frequency domain, and to enhance their detection
sensitivity. The electrostatic force can be modulated and nullified to obtain
surface potential maps (Kelvin probe force microscopy,
KFM)\cite{Nonnenmacher:1991ct,Kitamura:1998jk}.

The theory of dynamic AFM describes the effects of external forces on the motion
of the AFM tip mounted on a cantilever. While the theory of specific detection
schemes, such as AM-AFM\cite{Martin:1987ht,Paulo:2002js} and
FM-AFM\cite{Albrecht:1991bu,Giessibl:1997cd,Sader:2005bc}, is well established,
a general theory of steady-state operation with a minimum number of assumptions
was only developed recently\cite{Songen:2017kc}.

In this paper, we describe a unified theory of dynamic AFM that extends beyond
the steady-state solution by also considering the transient response to
perturbations.

First, we focus on the steady-state solution for excitation frequencies far
below and at resonance of the cantilever. We approach this problem from the
frequency domain, allowing us to make the simple connection from Fourier
components of the tip--sample force to the cantilever response at different
harmonics.

For excitation close to resonance, we present analytical expressions under the
so-called \emph{harmonic approximation}\cite{Songen:2017kc}, which assumes a
purely sinusoidal cantilever movement and thereby disregards harmonics caused by
the non-linear tip--sample force. As discussed later, this approximation is
justified for most force sensors and operating conditions. We demonstrate the
generality of the obtained steady-state solution to explain the operational
bistabilty in AM-AFM\cite{Nony:1999bp,Garcia:1999dm,Garcia:2000ij}.

Next, we go beyond the steady state and investigate the transient
cantilever response after perturbations of the tip--sample force or the
sinusoidal excitation. The behavior of transients is derived analytically in the
Laplace domain, which facilitates straightforward interpretation in the
frequency domain and enables the calculation of the response to an
arbitrarily-shaped perturbation. We show how perturbations of the external force
at different frequencies affect the tip movement, and how frequency mixing due
to the non-linear tip--sample interaction can be exploited.

The transients are then used to derive the baseband dynamics of the tip, which
are experimentally accessible by coherent demodulation of the cantilever
deflection at the excitation frequency. The complex baseband signal contains
information about the amplitude and phase modulation caused by the tip--sample
force. Through the baseband signal, we derive transfer functions from frequency
components of the external force to the amplitude, phase, and frequency response
of the cantilever.

In practice, these transfer functions are important for correct tuning of
feedback loops and signal-to-noise analysis of advanced AFM techniques, such as
multi-frequency\cite{Garcia:2012gq}, heterodyne\cite{Sugawara:2012bh} and
multi-harmonic\cite{Garrett:2018ck} modes of operation.

We use the derived transfer functions to determine the
actual closed-loop behavior of the PLL and amplitude controller in FM-AFM. The
behavior of the PLL is discussed for different controller gains and quality
factors of the cantilever.

Lastly, we highlight how the theory presented in this paper benefits AFM
simulators. The separation of the calculation into steady-state and transients
permits a large speedup compared to computationally expensive time-domain
simulations of the full equation of motion. The simulation can be done directly
at the baseband level, which mitigates the calculation of high-frequency
perturbations that are filtered out for common operational modes.

\section{Equation of motion}

The motion of the tip of an atomic force microscope is well approximated by the
behavior of a damped harmonic oscillator\cite{Garcia:2002kc}. In the vicinity of
the surface, the tip--sample force \(F_{\mathrm{ts}}\) acts on the tip at the
position \(z_{\mathrm{t}}\)
\begin{align}
    \ddot{z}_{\mathrm{t}} + \frac{\omega_0}{Q} \dot{z}_{\mathrm{t}}
    + \omega_0^2 \left( z_{\mathrm{t}} - z_{\mathrm{b}} \right)
    =
    \frac{\omega_0^2}{k} F_{\mathrm{ts}}(z_{\mathrm{ts}}, \dot{z}_{\mathrm{ts}}, t),
    \label{eq:eq-of-motion-tip}
\end{align}
where the eigenfrequency \(\omega_0\) and intrinsic dissipation, expressed as
the quality factor \(Q\), determine the shape of the resonance, and \(k\) is the
spring constant of the cantilever. \(z_{\mathrm{b}}\) is base position of the
cantilever and \(z_{\mathrm{ts}}\) is the tip--sample distance.

Experimentally, the deflection \(q = z_{\mathrm{t}} - z_{\mathrm{b}}\) of the
tip rather than its position \(z_{\mathrm{t}}\) is measured. The corresponding
equation of motion is therefore
\begin{align}
    \ddot{q} + \frac{\omega_0}{Q} \dot{q} + \omega_0^2 q
    &=
    \omega_0^2 a(t)
    + \frac{\omega_0^2}{k} F_{\mathrm{ts}}(z_{\mathrm{ts}}, \dot{z}_{\mathrm{ts}}, t),
    \label{eq:eq-of-motion}
\end{align}
where \(a(t)\) is the excitation of the tip resulting from the movement of the
base\cite{Platz:2012ey}:
\begin{align}
a(t) &= -\frac{\ddot{z}_{\mathrm{b}}}{\omega_0^2} - \frac{\dot{z}_{\mathrm{b}}}{Q \omega_0}.
\label{eq:excitation-base}
\end{align}
For a sinusoidal displacement of the base at the frequency \(\omega\),
\cref{eq:excitation-base} corresponds to a static change in amplitude and phase
by \((\omega / \omega_0)^2\) and \(\arctan(-\omega_0 / \omega Q)\),
respectively. In the limit of low intrinsic damping and drive near resonance
(\(\omega \approx \omega_0, Q \rightarrow \infty\)), the amplitude and phase are
left unchanged, and there is no difference exciting the base or
tip\cite{Garbini:1996fk,Platz:2012ey}. For the sake of clarity, we consider in
the following a direct excitation of the tip via \(a(t)\) rather than a
movement of the base. This has no implication on the generality of the results;
however, the effects of \cref{eq:excitation-base} must be considered separately
when operating off resonance or in an environment of high damping.

\section{Steady-state solution}
\label{sec:steadystate}

In the steady-state, transients due to the explicit time dependency of
\(F_{\mathrm{ts}}\) must have decayed. \(F_{\mathrm{ts}}(z_{\mathrm{ts}}(t),
\dot{z}_{\mathrm{ts}}(t))\) is, in general, a non-linear function of the tip
trajectory. Hence, for sinusoidal excitation of the oscillator at the frequency
\(\omega\) with an amplitude \(a_0\),
\begin{align}
    a(t) &= a_0 \cos(\omega t),
    \label{eq:excitation}
\end{align}
a response of the deflection \(q(t)\) at multiple harmonics (\(\omega, 2\omega,
3\omega, \dots\)) is expected.

Because the excitation signal \(a(t) = a(t+T)\) is periodic in time with the
period \(T=2\pi/\omega\), the tip trajectory \(z_{\mathrm{ts}}\) and the
tip--sample force \(F_{\mathrm{ts}}\) must also be periodic functions with the
same period \(T\). Therefore, \(q\) and \(F_{\mathrm{ts}}\) can be written as a
Fourier series,
\begin{align}
    q(t) = \sum_{n = -\infty}^{\infty} \hat{q}_n e^{i n \omega t}
    \quad \text{and} \quad
    F_{\mathrm{ts}}(t) = \sum_{n = -\infty}^{\infty} \hat{f}_n e^{i n \omega t}
    \label{eq:fourier-ansatz}
\end{align}

Inserting \cref{eq:excitation} and \cref{eq:fourier-ansatz} in
\cref{eq:eq-of-motion}, we obtain a system of equations for the Fourier
coefficients of the deflection \(\hat{q}_n\),
\begin{align}
    k \hat{q}_0 &= \hat{f}_0, \label{eq:fourier-static-deflection} \\
    k \hat{q}_{\pm 1} \left[
        -\omega^2
        \pm i \frac{\omega \omega_0}{Q}
        + \omega_0^2
        \right]
    &=
    k \omega_0^2 \frac{a_0}{2} + \omega_0^2 \hat{f}_{\pm 1},
    \label{eq:fourier-fundamental-harmonic} \\
    k \hat{q}_{\pm n} \left[
        -(\pm n \omega)^2
        \pm i \frac{n \omega \omega_0}{Q}
        + \omega_0^2
        \right]
    &=
    \omega_0^2 \hat{f}_{\pm n}.
    \label{eq:fourier-higher-harmonics}
\end{align}

\Cref{eq:fourier-static-deflection} describes the static deflection of the
cantilever, which is related to \(\hat{f}_0\) by the spring constant \(k\).
\Cref{eq:fourier-fundamental-harmonic} describes the response of the fundamental
harmonic of the deflection, \(\hat{q}_{\pm 1}\). This response is driven both by
the excitation force \(k a_0\) at the drive frequency \(\omega\), and the
Fourier component \(\hat{f}_{\pm 1}\). The Fourier coefficients for higher
harmonics \(|n| > 1\) follow from \cref{eq:fourier-higher-harmonics}.

For the solution of
\cref{eq:fourier-static-deflection,eq:fourier-fundamental-harmonic,eq:fourier-higher-harmonics},
the tip trajectory obtained from the Fourier components \(\hat{q}_n\) of the deflection,
is required to calculate the Fourier coefficients \(\hat{f}_n\)
of \(F_{\mathrm{ts}}(z_{\mathrm{ts}}(t), \dot{z}_{\mathrm{ts}}(t))\); likewise,
\(\hat{q}_n\) requires all \(\hat{f}_n\) to be known. Therefore, in general, a
self-consistent approach must be employed to solve for \(\hat{q}_n\). An
efficient numerical solution is possible in many circumstances by Newton's
method. Since \(F_{\mathrm{ts}}\) is a non-linear function of
\(z_{\mathrm{ts}}\), there can be multiple solutions, and the resulting
deflection \(q\) may sensibly depend on the initial conditions. For the
numerical solution, special care must be taken when the tip movement becomes
unstable upon approach and retract. Experimentally, this is the case when
`snap-into-contact' and `pull-off' events are observed\cite{Platz:2012ey}, i.e.,
when the force gradient \(k_{\mathrm{ts}}\) acting on the tip exceeds the spring
constant \(k\) of the cantilever. In such cases, the solution of
\cref{eq:fourier-static-deflection,eq:fourier-fundamental-harmonic,eq:fourier-higher-harmonics}
in the frequency domain can be cumbersome, whereas a time-domain solution of the
equation of motion, \cref{eq:eq-of-motion}, is usually straightforward. However,
as shown below, an analytical solution is possible when the fundamental harmonic
dominates the cantilever movement.

The number of significant harmonics observed in the cantilever deflection
depends on the relative drive frequency \(\omega / \omega_0\). For an oscillator
driven near resonance, i.e., \(\omega / \omega_0 \approx 1\), harmonics of the
drive frequency are rarely observed. This can be seen from
\cref{eq:fourier-fundamental-harmonic,eq:fourier-higher-harmonics}: to obtain
each \(\hat{q}_n\), \(\hat{f}_n\) is divided by a gain factor on the left hand
side. For \(\omega \approx \omega_0\), the gain ratio of the fundamental to the
\(n\)-th harmonic is \(\sqrt{Q^2 (1-n^2)^2 + n^2}\). Hence, the harmonics at \(n
\omega_0\) are off-resonance and diminish with increasing order \(n\) and
quality factor \(Q\). For reasonable quality factors \(Q\) exceeding the order
\(n\), i.e.\ \(Q \gg n\), harmonics decrease in amplitude as \(1 / Q (n^2-1)\).
Even in the limit of a highly overdamped oscillator, \(Q \rightarrow 0\),
harmonics still decrease with increasing order as \(1/n\). For harmonics to be
relevant and within the detection limits, the oscillation amplitude must be
sufficiently high to generate strong harmonics, and the quality factor must be
very low\cite{Preiner:2007bs,Kuchuk:2015jf}.

When the oscillator is instead driven off-resonance at frequencies \(\omega\)
well below the eigenfrequency \(\omega_0\), the \(n\)-th harmonic is attenuated
for \(n\omega \gg \omega_0\) and amplified near resonance. This is the case for
AFM techniques based on fast acquisition of force--distance
curves\cite{RosaZeiser:1997kx,Amo:2016ck}.
In this case, the reconstruction of the tip--sample force during each
oscillation cycle is possible via
\cref{eq:fourier-static-deflection,eq:fourier-fundamental-harmonic,eq:fourier-higher-harmonics}
from the deflection \(\hat{q}_n\) and the parameters of the freely oscillating
cantilever. In doing so, care must be taken to avoid amplifying noise
for higher harmonics beyond \(\omega_0\).

Alternatively, the non-linear tip--sample force can be reconstructed from its
harmonics by driving the cantilever at more than one frequency near the
resonance. Because of frequency mixing, intermodulation products appear in the
deflection signal\cite{Platz:2008ic,Platz:2012ey}. By appropriate choice of the
driving frequencies and their spacing, many intermodulation products are within
the resonance, such that they can be detected with a high signal-to-noise ratio.

In the following, we consider the case where the fundamental harmonic dominates
the deflection signal. As reasoned above, this \emph{harmonic
approximation} is valid in most experiments with a
single-tone excitation near the cantilever resonance. Experimentally, the
validity of this approximation can be verfied by observing the power spectrum of
\(q(t)\) for harmonics of \(\omega\).

Under the harmonic approximation, the deflection and tip--sample distance are
\begin{align}
    q &= q_{\mathrm{s}} + A \cos(\omega t + \phi) \label{eq:q-steadystate}\\
    z_{\mathrm{ts}} &= z_{\mathrm{b}} + q = z_{\mathrm{c}} + A \cos(\omega t + \phi),
    \label{eq:zts-steadystate}
\end{align}
where \(q_{\mathrm{s}}\) is the static deflection, \(z_{\mathrm{c}} = z_{\mathrm{b}} +
q_{\mathrm{s}}\) is the average (center) tip distance, and \(A\) and \(\phi\)
are the oscillation amplitude and phase, respectively.

The static deflection is governed by the time average of the tip--sample force,
\begin{align}
    \hat{f}_0
        &= \frac{1}{T} \int_{-T/2}^{T/2} \mathrm{d}t \,
        F_{\mathrm{ts}}(z_{\mathrm{ts}}(t), \dot{z}_{\mathrm{ts}}(t)) \nonumber\\
        &= \frac{1}{\pi} \int_{-A}^{A} \mathrm{d}q \,
        \frac{F_{\mathrm{ts}}(z_{\mathrm{c}} + q)}{\sqrt{A^2-q^2}}
        =: \langle F_{\mathrm{ts}} \rangle.
        \label{eq:Fts-avg}
\end{align}

The steady-state solution, \cref{eq:q-steadystate}, requires
\begin{align}
    \hat{q}_0 = q_{\mathrm{s}}
    \quad \text{ and } \quad
    \hat{q}_{\pm 1} = A \exp(\pm i\phi)/2.
    \label{eq:q-fourier}
\end{align}

For nonzero amplitudes \(A\), we can always write \(\hat{f}_{\pm 1}\) as a
product of a complex number \(\hat{k}_{\pm 1}\) and \(\hat{q}_{\pm 1}\),
corresponding to scaling and rotation of the complex force gradient
\(\hat{k}_{\pm 1}\):
\begin{align}
    \hat{f}_{\pm 1} = \hat{k}_{\pm 1} \hat{q}_{\pm 1}.
    \label{eq:f-fourier-decomp}
\end{align}

Because \(q(t)\) and \(F_{\mathrm{ts}}(t)\) are real signals, we can restrict the further
analysis to non-negative Fourier coefficients.
From the left side of \cref{eq:fourier-fundamental-harmonic}, we see that the
real and imaginary components of \(\hat{k}_{1}\) change the eigenfrequency
and damping to effective new values:
\begin{align}
    \omega_0^2 &\rightarrow
        \omega_0^2 \left( 1 - \operatorname{Re} \hat{k}_{1} / k \right)
        \label{eq:w0sq-eff}\\
    \frac{1}{Q} &\rightarrow
        \frac{1}{Q} - \frac{\omega_0}{\omega} \operatorname{Im} \hat{k}_{1} / k
        \label{eq:invQ-eff}
\end{align}

\(\hat{k}_{1}\) is related to the tip--sample force as
\begin{align}
    \hat{k}_{1} &=
        \frac{2 \hat{f}_{1}}{A e^{i \phi}}
        =
        \frac{2}{A T} \int_{t_0}^{t_0 + T} \mathrm{d}t
        F_{\mathrm{ts}}(t) e^{- i (\omega t + \phi)}.
\end{align}

For the real and imaginary parts of \(\hat{k}_1\), we find
\begin{align}
    \operatorname{Re} \hat{k}_1 &=
        \frac{2}{A T} \int_{t_0}^{t_0 + T} \mathrm{d}t \,
        F_{\mathrm{ts}}(t) \cos(\omega t + \phi) \label{eq:re-k1}\\
    \operatorname{Im} \hat{k}_1 &=
        -\frac{2}{A T} \int_{t_0}^{t_0 + T} \mathrm{d}t \,
        F_{\mathrm{ts}}(t) \sin(\omega t + \phi) \label{eq:re-k2}
\end{align}

\Cref{eq:re-k1,eq:re-k2} suggest that changes of eigenfrequency and damping are
determined by the in-phase and quadrature components of \(F_{\mathrm{ts}}\) with
respect to the tip oscillation. The integrals in \cref{eq:re-k1,eq:re-k2} are
zero if \(F_{\mathrm{ts}}(t-\phi/\omega)\) is odd or even in \(t\),
respectively. As suggested by \citet{Sader:2005bc}, this motivates splitting
\(F_{\mathrm{ts}}\) into even and odd parts, which govern changes of the
eigenfrequency and damping, respectively.

The odd force can be written in terms of a product of an even and odd function.
Naturally, choosing the tip velocity as the odd function, this results in the
generalized damping coefficient\cite{Holscher:2001kx,Sader:2005bc}
\(\gamma_{\mathrm{ts}}\),
\begin{align}
    F_{\mathrm{odd}}(z_{\mathrm{ts}}, \dot{z}_{\mathrm{ts}})
        &= -\gamma_{\mathrm{ts}}(z_{\mathrm{ts}}) \dot{z}_{\mathrm{ts}}.
\end{align}

Using the even and odd components \(F_{\mathrm{even}}\) and
\(F_{\mathrm{odd}}\) of \(F_{\mathrm{ts}}\), and substituting the tip trajectory
\(z_{\mathrm{ts}}(t)\), we obtain
\begin{align}
    \operatorname{Re} \hat{k}_1
        &= \frac{2}{\pi A^2} \int_{-A}^{A} \mathrm{d}q \,
        \frac{F_{\mathrm{even}}(z_{\mathrm{c}} + q) \, q}{\sqrt{A^2-q^2}} \nonumber\\
        &= \frac{2}{\pi A^2} \int_{-A}^{A} \mathrm{d}q \,
        k_{\mathrm{ts}}(z_{\mathrm{c}} + q) \, \sqrt{A^2-q^2} \nonumber\\
        &=: \langle k_{\mathrm{ts}} \rangle
        \label{eq:kts-avg} \quad \text{and}\\
    \operatorname{Im} \hat{k}_1
        &= -\omega \frac{2}{\pi A^2} \int_{-A}^{A} \mathrm{d}q \,
        \gamma_{\mathrm{ts}}(z_{\mathrm{c}} + q) \, \sqrt{A^2-q^2} \nonumber\\
        &=: -\omega \langle \gamma_{\mathrm{ts}} \rangle,
        \label{eq:gamma-avg}
\end{align}
with the force gradient \(k_{\mathrm{ts}} = \partial F_{\mathrm{even}}/\partial
z\), and \(\langle k_{\mathrm{ts}} \rangle\) and  \(\langle \gamma_{\mathrm{ts}}
\rangle\) are the effective force gradient and damping coefficient,
respectively. Unlike \(\langle F_{\mathrm{ts}} \rangle\), which represents a
simple time average of the force and whose main contributions are due to the
turning points of the oscillation, the time averages for \(\langle
k_{\mathrm{ts}} \rangle\) and \(\langle \gamma_{\mathrm{ts}} \rangle\) are
weighted by the oscillation itself, which results in a weighting over a
semi-circle around \(z_{\mathrm{c}}\).

The formula for \(\langle k_{\mathrm{ts}} \rangle\) was first derived by
\citet{Giessibl:1997cd} using a Hamilton-Jacobi perturbation approach and later
rewritten as a weighted average over \(k_{\mathrm{ts}}\) by integration by
parts\cite{Giessibl:2001ig}. A similar Fourier \emph{ansatz} to the one shown
here has been employed before for
FM-AFM\cite{Holscher:2001kx,Sader:2005bc,Ebeling:2007dz} and
AM-AFM\cite{Holscher:2007fb}.

Recently, from a consideration of the average kinetic energy and average power
of the resonator, \citet{Songen:2017kc} derived three equations relating the
excitation parameters (\(a_0\),
\(\omega\)) and observables (\(q_{\mathrm{s}}\), \(A\), \(\phi\)) of
typical dynamic AFM experiments to the averages
\(\langle F_{\mathrm{ts}} \rangle\), \(\langle k_{\mathrm{ts}} \rangle\), and
\(\langle \gamma_{\mathrm{ts}} \rangle\).
These equations also follow naturally from
\cref{eq:fourier-static-deflection,eq:fourier-fundamental-harmonic}. Inserting
the observables from \cref{eq:q-fourier} and using the averages defined in
\cref{eq:Fts-avg,eq:kts-avg,eq:gamma-avg}, we obtain:
\begin{subequations}
\begin{align}
    \langle F_{\mathrm{ts}} \rangle / k &= q_{\mathrm{s}}
        \label{eq:afm-Fts} \\
    \langle k_{\mathrm{ts}} \rangle / k
        &= 1 - \left(\omega / \omega_0\right)^2
        - \frac{a_0}{A} \cos \phi \label{eq:afm-kts} \\
        &\approx -2 \frac{\Delta \omega}{\omega_0} - \frac{a_0}{A} \cos \phi
        \label{eq:kts-k-approx}
        \\
    \langle \gamma_{\mathrm{ts}} \rangle / k
        &= -\frac{1}{\omega_0 Q} - \frac{a_0}{\omega A} \sin \phi.
        \label{eq:afm-gamma}
\end{align}
\end{subequations}
In \cref{eq:kts-k-approx} we have used \(\omega = \omega_0 + \Delta \omega\) and
approximated for small frequency shifts \(\Delta \omega\). The equations are
useful to quantitatively compare measurements obtained by different AFM
implementations, as demonstrated for AM- and FM-AFM spectroscopy
data\cite{Songen:2017kc}. It should be noted that the averages themselves are
non-linear functions of distance \(z_{\mathrm{c}}\) and amplitude \(A\); a
direct comparison is therefore only possible for data collected with a similar
amplitude, or after deconvolving the averages to reconstruct the tip--sample
force\cite{Songen:2017kc,Sader:2004kt,Sader:2005bc}.

In similar form, \cref{eq:afm-kts,eq:afm-gamma} have been derived before by
\citet{Paulo:2002js} to obtain a general theory of AM-AFM. The relation of
dissipation and oscillation phase, \cref{eq:afm-gamma}, was first given by
\citet{Cleveland:1998cn}.

Eliminating the phase in \cref{eq:afm-kts,eq:afm-gamma}, we obtain an algebraic
equation for the oscillation amplitude \(A\) as a function of excitation amplitude
\(a_0\) and frequency \(\omega\):
\begin{align}
    \frac{a_0}{A}
    =
    \sqrt{
    \bigg[ 1 - \bigg( \frac{\omega}{\omega_0} \bigg)^2 - \frac{\langle k_{\mathrm{ts}} \rangle}{k} \bigg]^2
    + \bigg[ \frac{\omega}{\omega_0 Q} + \frac{\omega \langle \gamma_{\mathrm{ts}} \rangle}{k} \bigg]^2
    }
    \label{eq:excitation-to-amplitude}
\end{align}
For fixed excitation parameters \(a_0\) and \(\omega\), the solution of
\cref{eq:excitation-to-amplitude} are the oscillation amplitudes \(A\) allowed
at the distance \(z_{\mathrm{c}}\).

The time-averaged power dissipated by the tip,
\(\langle P_{\mathrm{tip}} \rangle\), is related to the
driving frequency \(\omega\), steady-state amplitude \(A\), and the tip--sample
damping coefficient \(\langle \gamma_{\mathrm{ts}} \rangle\):
\begin{align}
    \langle P_{\mathrm{tip}} \rangle
        &= -\langle F_{\mathrm{odd}} \dot{z}_{\mathrm{ts}} \rangle
        = \langle \gamma_{\mathrm{ts}} \dot{z}_{\mathrm{ts}}^2 \rangle
        = \omega^2 A^2 \langle \gamma_{\mathrm{ts}} \rangle / 2.
\end{align}

Motivated by the substitution in \cref{eq:w0sq-eff,eq:invQ-eff}, the original
non-linear tip--sample force \(F_{\mathrm{ts}}\) can be approximated in terms of the
averages \(\langle F_{\mathrm{ts}} \rangle\), \(\langle k_{\mathrm{ts}}
\rangle\), and \(\langle \gamma_{\mathrm{ts}} \rangle\). Therefore,
\(F_{\mathrm{ts}}\) in \cref{eq:eq-of-motion} may be substituted with
\begin{align}
    F_{\mathrm{ts}} &\approx
        \langle F_{\mathrm{ts}} \rangle (t)
        + \langle k_{\mathrm{ts}} \rangle (t) \, (q - q_{\mathrm{s}})
        - \langle \gamma_{\mathrm{ts}} \rangle (t) \, \dot{q}.
    \label{eq:Fts-harmonic-approx}
\end{align}
Note that the individual terms may still be treated as weak functions of time,
since the averages in \cref{eq:Fts-avg,eq:kts-avg,eq:gamma-avg} can be taken as
short as a single period of oscillation. \Cref{eq:Fts-harmonic-approx} may be
considered an equivalent linearization\cite{Mitropolskii:1997dr} of the
non-linear equation of motion. The solution derived from
the harmonic approximation is therefore equivalent to the one obtained using the
Krylov-Bogoliubov averaging method\cite{Krylov:1949tc,Mitropolskii:1997dr}.
\Cref{eq:Fts-harmonic-approx} approximates the tip--sample force by its Fourier
coefficients \(\hat{f}_0\) and \(\hat{f}_{\pm1}\) only.

\begin{figure*}
    \includegraphics[width=\linewidth]{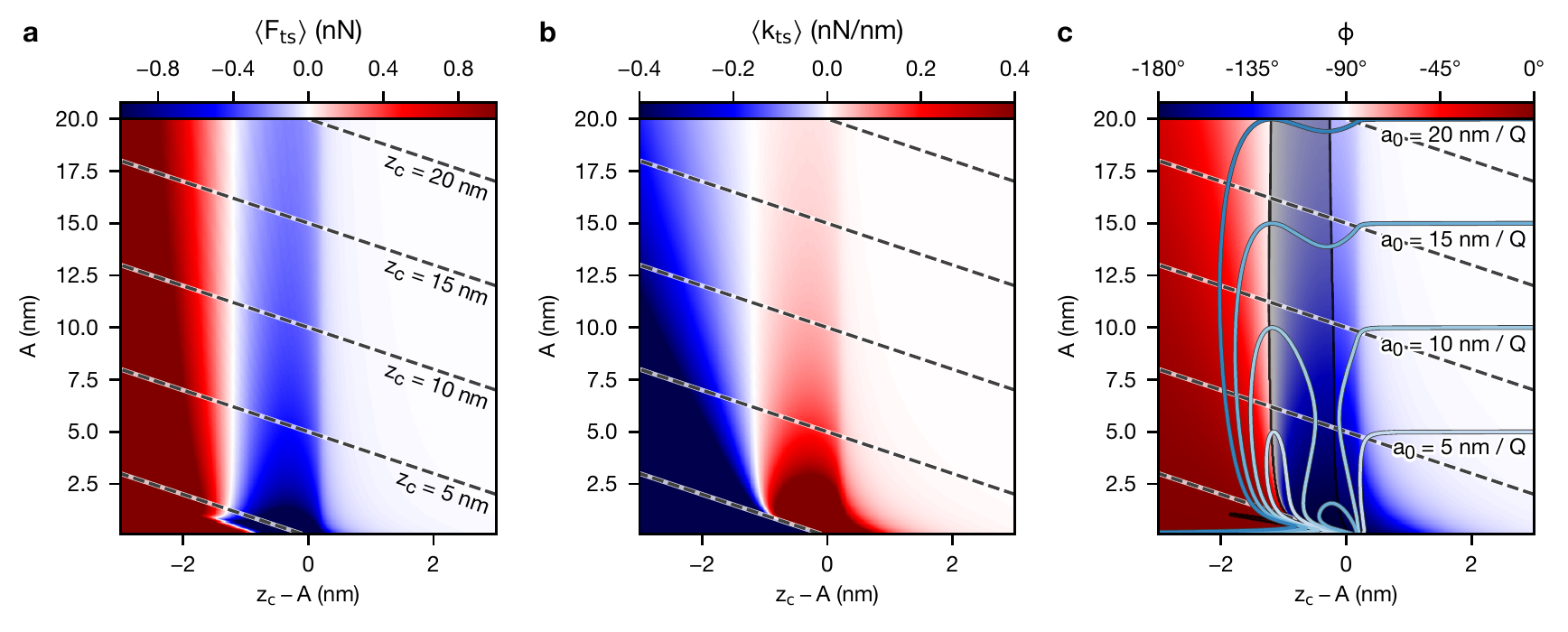}
    \caption{
    Effective (a)~tip--sample force and (b)~force gradient as a function of
    amplitude \(A\) and distance of closest approach \(z_{\mathrm{c}} - A\).
    Tip--sample interactions are given by a vdW-DMT model as parameterized in
    \citet{Paulo:2002js} (\(H = \SI{6.4e-20}{J}\), \(E^\ast = \SI{1}{GPa}\),
    \(c = \SI{0.17}{nm}\), \(R =
    \SI{10}{nm}\)). Dashed lines indicate contours of constant average
    tip--sample distance \(z_{\mathrm{c}}\).
    (c) Phase and contours of constant excitation resulting from
    \cref{eq:afm-kts,eq:excitation-to-amplitude}
    for a cantilever with \(\omega_0 = \omega = 2\pi \, \SI{325}{kHz}\), \(Q =
    400\), and \(k = \SI{40}{\newton\per\meter}\).
    Note that for constant excitation \(a_0\), a steady-state oscillation with a
    certain amplitude \(A\) is possible at multiple distances. The region shaded
    in gray indicates solutions which are unstable, that is, small
    perturbations result in a jump to a nearby stable solution.
    }
    \label{fig:harmonic-solution}
\end{figure*}

As an application of the steady-state solution discussed above, we show in
\cref{fig:harmonic-solution} the averages \(\langle F_{\mathrm{ts}}
\rangle\) and \(\langle k_{\mathrm{ts}} \rangle\) as function of amplitude \(A\)
and distance of closest approach, \(z_{\mathrm{c}}-A\), together with the
phase and excitation calculated from \cref{eq:afm-kts,eq:afm-gamma}. Tip--sample
interactions are given by a Derjaguin-Muller-Toporov (DMT) contact model
combined with van der Waals (vdW) interactions, which is commonly employed in
literature\cite{Garcia:1999dm,Garcia:2002kc,Paulo:2002js,Platz:2012ey}:
\begin{align}
    F_{\mathrm{ts}}(z) &=
    \begin{cases}
        -\frac{H R}{6 z^2} & z > c \\
        -\frac{H R}{6 c^2} + \frac{4}{3} E^\ast \sqrt{R (c - z)^3} & z \leq c
    \end{cases}
\end{align}
Here, \(H\) is the Hamaker constant, \(E^\ast\) is the effective Young's modulus
of the tip--sample system, and \(c\) is the distance at which the tip leaves the
contact regime.

Notably, \cref{fig:harmonic-solution}(c) shows the bistability commonly observed
under AM-AFM operation\cite{Garcia:1999dm,Garcia:2000ij,Paulo:2002js}: for a
cantilever which is driven at fixed excitation \(a_0\), there can be several
possible distances \(z_{\mathrm{c}}\) resulting in an oscillation with the
amplitude \(A\).

Not every combination of \(A\) and \(z_{\mathrm{c}}\) is stable, however. For
the gray shaded area in \cref{fig:harmonic-solution}(c), a small perturbation
is sufficient to drive the oscillation to the nearest stable branch. This region
was found by a numerical simulation of the equation of motion,
\cref{eq:eq-of-motion}, together with the effective \(F_{\mathrm{ts}}\),
\cref{eq:Fts-harmonic-approx}. The averages are calculated efficiently using
Chebyshev--Gauss quadrature and are precomputed on a grid as shown in
\cref{fig:harmonic-solution}(a) and (b). The equation of motion is evaluated at
the baseband (c.f.\ \cref{sec:baseband}), such that the resulting amplitude and
phase are available without further signal processing.

For large free amplitudes (\(> \SI{10}{nm}\)), net-repulsive tip--sample
interactions are favored, because there exists no net-attractive branch at
typical imaging setpoints of \(80-\SI{90}{\percent}\). For smaller amplitudes
(\(< \SI{10}{nm}\)), there are both stable net-attractive and stable
net-repulsive branches for most amplitudes below the free amplitude. Operation
in the net-attractive regime is favored increasingly, because the separation of
the stable branches is increased, and strong perturbations are required to
facilitate a jump between them.

\section{Transients}
\label{sec:transients}

Next, we discuss the response of the tip deflection \(q(t)\) to changes of the
external forces. To this end, it is instructive to work with the Laplace
transform \(\tilde{q}(s)=\mathcal{L}\{q\}(s)\) instead, because the time-domain
behavior is then reflected in algebraic expressions of the complex variable
\(s=\sigma+i\omega\).

Far from the surface, the cantilever behavior is found from the Laplace transform
of \cref{eq:eq-of-motion} with \(F_{\mathrm{ts}} \equiv 0\),
\begin{align}
    \tilde{q}(s) &= G_0(s) \tilde{a}(s) \quad \text{with} \\
    G_0(s) &= \frac{\omega_0^2}{s^2 + \omega_0 s / Q + \omega_0^2},
    \label{eq:transfer-function-cantilever}
\end{align}
where we denote \(\tilde{q}(s)\) and \(\tilde{a}(s)\) as the Laplace-transformed
deflection and drive amplitude, respectively. \(G_0(s)\) is the well-known
transfer function of a harmonic oscillator.
By partial fraction expansion, this transfer function can be rewritten in terms
of the complex conjugate poles \(p\) and \(p^*\) of the denominator,
\begin{align}
    G_0(s) &= \frac{\omega_0^2}{p - p^*}
    \left(
        \frac{1}{s-p} - \frac{1}{s-p^*}
    \right),
    \label{eq:transfer-function-cantilever-expansion} \\
    \text{with} &\quad
    p = -\omega_{\mathrm{c}} + i \sqrt{\omega_0^2 - \omega_{\mathrm{c}}^2}
    \;\;\text{and}\;\; \omega_{\mathrm{c}} = \omega_0 / 2Q.
    \label{eq:transfer-function-cantilever-expansion-poles}
\end{align}
Here, \(\omega_{\mathrm{c}}\) is introduced as a cutoff frequency, which is
commonly known as the cantilever bandwidth. The transfer function written as
\cref{eq:transfer-function-cantilever-expansion} is particularly useful, because
it separates the second-order system into the equivalent of two parallel
first-order systems.
\begin{figure}
    \includegraphics[width=\linewidth]{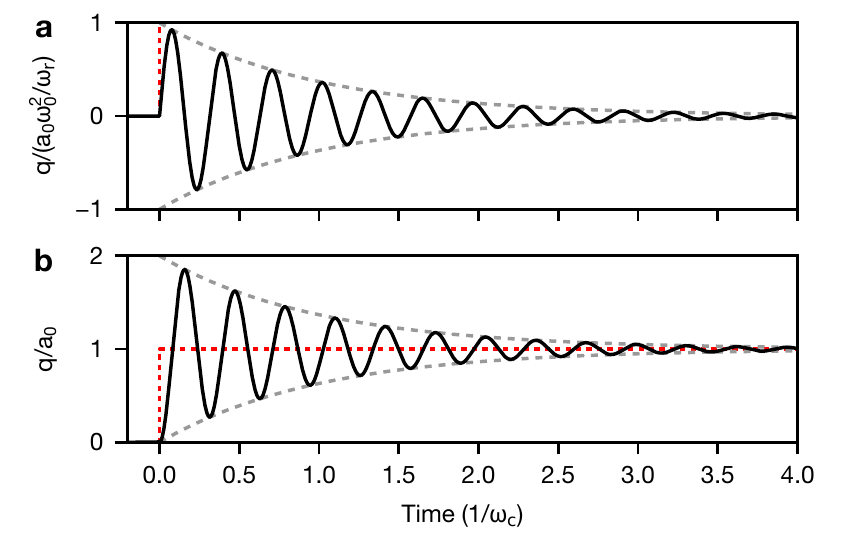}
    \caption{
        Response of the deflection \(q(t)\) to an impulse~(a) and step~(b) of
        the excitation \(a(t)\) at \(t=0\) for a harmonic oscillator with
        \(\omega_0=2\pi \, \SI{100}{kHz}\) and \(Q=10\). The gray dashed lines
        indicate the envelope of the decaying oscillations at the resonance
        frequency \(\omega_\mathrm{r} \approx \omega_0\).
    }
    \label{fig:transients}
\end{figure}
In the time-domain, the impulse response of each subsystem with a pole \(p\) can
readily be found by an inverse Laplace transform as \(\exp(p\,t)\), which
corresponds to an exponential decay \(\exp(-\omega_{\mathrm{c}} t)\) of an
oscillation excited at the resonance frequency \(\omega_\mathrm{r} =
\sqrt{\omega_0^2 - \omega_{\mathrm{c}}^2}\). The impulse and step responses
following \cref{eq:transfer-function-cantilever-expansion} are illustrated in
\cref{fig:transients}.

Near the surface, with \cref{eq:eq-of-motion,eq:Fts-harmonic-approx}, the
cantilever behavior around the steady-state solution is
\begin{align}
    \ddot{q}
    + \omega_0 \left( \frac{1}{Q} + \frac{\omega_0}{k} \langle \gamma_{\mathrm{ts}} \rangle \right) \dot{q}
    + \omega_0^2 \left( 1 - \frac{\langle k_{\mathrm{ts}} \rangle }{k} \right) q
    \nonumber\\
    = \omega_0^2 a(t)
    + \frac{\omega_0^2}{k} \delta F_{\mathrm{int}}(t).
    \label{eq:eq-of-motion-approx}
\end{align}

\(\delta F_{\mathrm{int}}\)(t) describes deviations of the tip--sample force from
the steady state. In general, \(\delta F_{\mathrm{int}}\) follows from the Fourier
series of \(F_{\mathrm{ts}}\), \cref{eq:fourier-ansatz}, as
\begin{align}
    \delta F_{\mathrm{int}}(t)
        &= \sum_{n=-\infty}^{\infty} \delta \hat{f}_n(t) e^{i n \omega t},
\end{align}
and the corresponding Laplace transform is
\begin{align}
    \delta \tilde{F}_{\mathrm{int}}(s)
        &= \sum_{n=-\infty}^{\infty} \delta \tilde{f}_n(s - i n \omega),
    \label{eq:delta-Fint}
\end{align}
where \(\delta\tilde{f}_n(s)\) corresponds to the Laplace transform of the
time-varying Fourier coefficient \(\delta\hat{f}_n(t)\).

Laplace transform of \cref{eq:eq-of-motion-approx} results in
\begin{align}
    \tilde{q}(s)
        &= G(s) \left( \tilde{a}(s) + \delta \tilde{F}_{\mathrm{int}} (s) / k \right),
        \label{eq:deflection-surface}
\end{align}
where \(G(s)\) is the steady-state transfer function of the cantilever, found by
replacing \(\omega_0^2\) and \(1/Q\) in the denominator of \(G_0(s)\) according
to \cref{eq:w0sq-eff,eq:invQ-eff}. Accordingly, due to the interaction with the
surface, the cantilever bandwidth and poles of \(G(s)\) are changed to
\begin{align}
    \omega_{\mathrm{c}} &\rightarrow
        \omega_{\mathrm{c}}' =
        \omega_{\mathrm{c}} + \langle \gamma_{\mathrm{ts}} \rangle \omega_0^2 / 2k
    \label{eq:surface-cantilever-bandwidth} \\
    p &\rightarrow
        p' =
        -\omega_{\mathrm{c}}'
        + i \sqrt{
            \omega_0^2 \left( 1 - \langle k_{\mathrm{ts}} \rangle / k \right)
            - \omega_{\mathrm{c}}^{\prime 2}
        }.
    \label{eq:surface-poles}
\end{align}

In general, we must also account for amplitude and phase changes of the
excitation signal introduced in \cref{eq:excitation}:
\begin{align}
    a(t) &= (a_0 + \delta a_{\mathrm{exc}}(t)) \cos(\omega t + \delta\phi_{\mathrm{exc}}(t)),
    \label{eq:excitation-perturbed}
\end{align}
where \(\delta a_{\mathrm{exc}}(t)\) and \(\delta\phi_{\mathrm{exc}}(t)\) denote
a small amplitude and phase modulation, respectively, which we add here to
find the effect of changes to the excitation over time. Assuming that
\(\delta\phi_{\mathrm{exc}}(t) \ll 1\), the Laplace transform of
\cref{eq:excitation-perturbed} can be written as
\begin{align}
    \tilde{a}(s) &= \tilde{a}_{\mathrm{s}}(s) + \delta \tilde{a}(s) \\
        &= \frac{a_0}{2}
            \Big( \frac{1}{s - i\omega} + \frac{1}{s + i\omega} \Big) \nonumber\\
        &\mathrel{\phantom{=}}
            + \frac{1}{2}
            \Big(
                \delta\tilde{a}_{\mathrm{exc}}(s - i\omega)
                + \delta\tilde{a}_{\mathrm{exc}}(s + i\omega)
            \Big) \nonumber\\
        &\mathrel{\phantom{=}}
            + \frac{a_0}{2}
            \Big(
                i \delta\tilde{\phi}_{\mathrm{exc}}(s - i\omega)
                - i \delta\tilde{\phi}_{\mathrm{exc}}(s + i\omega)
            \Big),
    \label{eq:laplace-excitation}
\end{align}
where \(\tilde{a}_{\mathrm{s}}\) denotes the steady-state excitation, and
\(\delta \tilde{a}\) is a perturbation of the excitation due to amplitude and
phase modulation.

The solution \(\tilde{q}(s)\) of \cref{eq:deflection-surface} can be separated
into the known solution of the steady state, \(\tilde{q}_{\mathrm{s}}(s)\), and
perturbations due to excitation \(\delta\tilde{q}_{\mathrm{exc}}(s)\) and
interaction \(\delta\tilde{q}_{\mathrm{int}}(s)\):
\begin{subequations}
\begin{align}
    \tilde{q}(s) &= \tilde{q}_{\mathrm{s}}(s)
        + \delta\tilde{q}_{\mathrm{exc}}(s)
        + \delta\tilde{q}_{\mathrm{int}}(s) \\
    \text{with}\quad
    \tilde{q}_{\mathrm{s}}(s) &= G(s) \, \tilde{a}_{\mathrm{s}}(s), \\
    \delta\tilde{q}_{\mathrm{exc}}(s) &= G(s) \, \delta \tilde{a}(s), \text{ and} \label{eq:delta-q-exc} \\
    \delta\tilde{q}_{\mathrm{int}}(s) &= G(s) \, \delta \tilde{F}_{\mathrm{int}} (s) / k. \label{eq:delta-q-int}
\end{align}
\end{subequations}

The steady-state perturbation of the deflection can be calculated
explicitly. Note that \(\tilde{a}_{\mathrm{s}}\) excites two oscillations at
\(\pm \omega\). The complex amplitudes of these oscillations are calculated
using the final value theorem:
\begin{align}
    \hat{q}_{\mathrm{s}}^{\pm} &:=
    \lim_{s \rightarrow 0} s \tilde{q}_{\mathrm{s}}(s \pm i\omega)
    = G(\pm i\omega) \frac{a_0}{2}.
\end{align}

Because the cantilever is driven by a real signal, its steady-state response
\(q_{\mathrm{s}}(t)\) must be real as well. \(\hat{q}_{\mathrm{s}}^{+}\) and
\(\hat{q}_{\mathrm{s}}^{-}\) must be complex conjugates:
\begin{align}
    \hat{q}_{\mathrm{s}}^+ =: \hat{q}_{\mathrm{s}} / 2
    \quad\text{and}\quad
    \hat{q}_{\mathrm{s}}^- = \hat{q}_{\mathrm{s}}^* / 2.
\end{align}

The steady-state contribution to the deflection is thus
\begin{align}
    \tilde{q}_{\mathrm{s}}(s)
        &= \frac{\hat{q}_{\mathrm{s}} / 2}{s - i\omega}
         + \frac{\hat{q}_{\mathrm{s}}^* / 2}{s + i\omega}.
    \label{eq:laplace-qs}
\end{align}

The perturbation of \(\tilde{q}(s)\) due to the excitation is
\begin{align}
    \delta\tilde{q}_{\mathrm{exc}}(s)
    &= \frac{1}{2} G(s) \, \Big[
        \delta\tilde{a}_{\mathrm{exc}}(s - i\omega)
        + \delta\tilde{a}_{\mathrm{exc}}(s + i\omega)
    \Big] \nonumber\\
    &\mathrel{\phantom{=}} + \frac{i a_0}{2} G(s) \, \Big[
        \delta\tilde{\phi}_{\mathrm{exc}}(s - i\omega)
        - \delta\tilde{\phi}_{\mathrm{exc}}(s + i\omega)
    \Big].
    \label{eq:delta-q-exc-explicit}
\end{align}

As a result of \cref{eq:Fts-avg,eq:f-fourier-decomp}, the dynamics of the zero-
 and first-order Fourier components of the transient interaction force \(\delta
 F_{\mathrm{int}}\) can be interpreted as perturbations
 of the tip--sample force and complex force gradient, respectively:
\begin{align}
    \delta \tilde{f}_0(s) &= \delta \tilde{F}_{\mathrm{ts}}(s) \\
    \delta \tilde{f}_{\pm 1}(s)
        &= \delta \tilde{k}_{\pm 1}(s) \, \hat{q}_{\mathrm{s}}^{\pm}
         = \left[ \delta \tilde{k}_{\mathrm{ts}}(s) \mp i \omega \tilde{\gamma}_{\mathrm{ts}}(s) \right]
           \hat{q}_{\mathrm{s}}^{\pm}.
\end{align}

Separating \(\hat{f}_0\) and \(\hat{f}_{\pm 1}\) from higher harmonics,
the transient interaction force, \cref{eq:delta-Fint}, can be written as
\begin{align}
    \delta F_{\mathrm{int}}(s)
    &= \delta \tilde{f}_0(s)
     + \delta \tilde{f}_1(s - i \omega)
     + \delta \tilde{f}_{-1}(s + i \omega) \nonumber\\
     &\mathrel{\phantom{=}}
     + \sum_{n=2}^{\infty}
        \left[
            \delta \tilde{f}_n(s - i n \omega) + \delta \tilde{f}_{-n}(s + i n \omega)
        \right].
    \label{eq:delta-Fint-explicit}
\end{align}

With \cref{eq:delta-q-int} the resulting perturbation of the deflection
\(\delta\tilde{q}_{\mathrm{int}}\) is therefore
\begin{align}
    \delta\tilde{q}_{\mathrm{int}}(s)
        &= \frac{1}{k} G(s) \bigg\{ \,
            \delta\tilde{F}_{\mathrm{ts}}(s)\nonumber\\
        &\hspace{0em} \phantom{\cdot \Big[} \, + \frac{1}{2} \Big[
                \delta\tilde{k}_{\mathrm{ts}}(s-i\omega)
                - i\omega \, \delta\tilde{\gamma}_{\mathrm{ts}}(s-i\omega)
            \Big] \hat{q}_{\mathrm{s}} \nonumber\\
        &\hspace{0em} \phantom{\cdot \Big[} \, + \frac{1}{2} \Big[
                \delta\tilde{k}_{\mathrm{ts}}(s+i\omega)
                + i\omega \, \delta\tilde{\gamma}_{\mathrm{ts}}(s+i\omega)
            \Big] \hat{q}_{\mathrm{s}}^* \nonumber\\
        &\hspace{0em} \phantom{\cdot \Big[} \, + \sum_{n=2}^{\infty}
        \left[
            \delta \tilde{f}_n(s - i n \omega) + \delta \tilde{f}_{-n}(s + i n \omega)
        \right]
        \bigg\}.
    \label{eq:delta-q-int-explicit}
\end{align}

This equation describes the behavior of the cantilever for arbitrarily chosen
perturbations of the force \(\delta\tilde{F}_{\mathrm{ts}}\), force gradient
\(\delta \tilde{k}_{\mathrm{ts}}\), damping coefficient \(\delta
\tilde{\gamma}_{\mathrm{ts}}\), or higher harmonics of the interaction force,
given only the approximation that the cantilever oscillation remains harmonic at
all times.
Perturbations of the tip--sample force \(\delta\tilde{F}_{\mathrm{ts}}\) drive
the cantilever directly via its steady-state transfer function \(G(s)\), whereas
perturbations of the force gradient \(\delta \tilde{k}_{\mathrm{ts}}\) or
damping coefficient \(\delta \tilde{\gamma}_{\mathrm{ts}}\) are shifted up and
down in the frequency domain by the drive frequency \(\omega\) before they enter
\(G(s)\). Similarly, contributions from the \(n\)-th harmonic of the interaction
force are frequency-shifted by \(\pm n \omega\).

\section{Baseband signal}
\label{sec:baseband}

The results of the previous section can be used to derive the dynamics of the
baseband signal, i.e., the signal detected by a lock-in amplifier at the
excitation frequency \(\omega\) and phase \(\delta\phi_{\mathrm{exc}}\).
In the time domain, the baseband signal is given by
\begin{align}
    q_{\mathrm{b}}(t)
        &= 2\, h(t) * q_\omega(t) \label{eq:baseband-time-domain} \\
        &\approx 2\, h(t) * \left[
            \left(1 - i \delta\phi_{\mathrm{exc}}(t)\right) e^{-i \omega t} q(t)
        \right], \label{eq:baseband-time-domain-approx}
\end{align}
where
\begin{align}
    q_\omega(t) &= \exp\left[ -i \omega t -i \delta\phi_{\mathrm{exc}}(t)\right] q(t)
\end{align}
denotes the down-conversion with respect to the excitation at \(\omega\), the
prefactor \(2\) is chosen to obtain peak amplitudes, \(h(t)\) is the impulse
response of a low-pass filter, and the asterisk \(*\) denotes a convolution. In
the following, we choose \(h(t)\) to filter frequencies beyond the modulation
frequency \(\omega\).

There are several ways to find \(q(t)\) and \(q_\omega(t)\) as required for the
solution. For example, \(q(t)\) can be obtained directly by solving the equation
of motion, \cref{eq:eq-of-motion}, e.g., by Verlet\cite{Verlet:1967cm}
integration or Runge-Kutta methods. Because we are interested in the demodulated
deflection \(q_\omega(t)\), we can make use of
\cref{eq:transfer-function-cantilever-expansion}, which describes the separation
of the second-order system into two first-order systems. After down-conversion,
\(\tilde{q}_\omega(s) = q(s + i \omega)\), one system describes the dynamics at
the baseband, while the other system contains high-frequency oscillations at
\(\approx \omega_0 + \omega\) which are removed by the low-pass filter.

With these considerations, the equation of motion for the baseband deflection
\(q_{\mathrm{b}}(t)\) is obtained by an inverse Laplace transform of
\cref{eq:transfer-function-cantilever-expansion} as
\begin{align}
    \dot{q}_{\mathrm{b}} - (p - i \omega) q_{\mathrm{b}} &= \frac{\omega_0^2}{p - p^*} a_{\mathrm{b}}(t),
    \label{eq:baseband-equation-of-motion}
\end{align}
where \(a_{\mathrm{b}}(t)\) is the baseband excitation signal. The poles \(p\)
and \(p^*\), given by \cref{eq:surface-poles}, include the interaction of the
tip with the surface through the averages \(\langle k_{\mathrm{ts}} \rangle\)
and \(\langle \gamma_{\mathrm{ts}} \rangle\).

Next, we derive the baseband dynamics in the frequency domain. Analytical
expressions and transfer functions are necessary for an intuitive understanding
of the cantilever behavior and for the design of feedback loops. Using the
Laplace transform of \cref{eq:baseband-time-domain-approx}, we obtain
\begin{align}
    \tilde{q}_{\mathrm{b}}(s)
        &\approx 2\, H(s) \left[
            \tilde{q}(s + i\omega)
            -i \delta\tilde{\phi}_{\mathrm{exc}}(s) * \tilde{q}(s + i\omega)
        \right] \\
        &\approx H(s) \left[
            2\, \tilde{q}(s + i\omega)
            -i \delta\tilde{\phi}_{\mathrm{exc}}(s) \hat{q}_{\mathrm{s}}
        \right],
    \label{eq:baseband-signal}
\end{align}
where \(H(s)\) indicates the transfer function of the low-pass filter. In the
latter step we use \(\tilde{q} \approx \tilde{q}_{\mathrm{s}}\) and assume that
\(\delta\tilde{\phi}\) vanishes for frequencies above \(\omega\).


In the following, we indicate filtered quantities by a prime, e.g., \(
\tilde{q}'(s) = H(s) \tilde{q}(s) \). The filtered contributions to the
deflection are
\begin{subequations}
\begin{align}
    2 \, \tilde{q}'_{\mathrm{s}}(s+i\omega)
        &= \hat{q}_{\mathrm{s}} / s,
    \label{eq:qs-downconv} \\
    2 \, \delta\tilde{q}'_{\mathrm{int}}(s+i\omega)
        &= \frac{1}{k} G(s + i\omega)
        \bigg\{
            \left[
                \delta\tilde{k}_{\mathrm{ts}}(s)
                - i\omega \delta\tilde{\gamma}_{\mathrm{ts}}(s)
            \right] \hat{q}_{\mathrm{s}} \nonumber \\
        &\phantom{=}
            + {\left[
                \delta\tilde{k}_{\mathrm{ts}}(s+2i\omega)
                + i\omega \delta\tilde{\gamma}_{\mathrm{ts}}(s+2i\omega)
            \right]} \hat{q}_{\mathrm{s}}^* \nonumber \\
        &\phantom{=}
            + 2 \, \delta\tilde{F}_{\mathrm{ts}}(s+i\omega)
        \bigg\}, \text{ and}
    \label{eq:delta-q-int-downconv} \\
    2 \, \delta\tilde{q}'_{\mathrm{exc}}(s+i\omega)
        &= {G(s+i\omega)}
        \left[
            \delta\tilde{a}_{\mathrm{exc}}(s)
            + \delta\tilde{\phi}_{\mathrm{exc}}(s) \, i a_0
        \right].
    \nonumber \\
    \label{eq:delta-q-exc-downconv}
\end{align}
\end{subequations}

For clarity of the presentation, higher harmonic terms of
\(\delta\tilde{q}_{\mathrm{int}}\) are left out in
\(\delta\tilde{q}'_{\mathrm{int}}\), because they are typically negligible if
they are not driven externally at a frequency of \((n \pm 1) \, \omega\).

Inserting \cref{eq:qs-downconv,eq:delta-q-int-downconv,eq:delta-q-exc-downconv}
into \cref{eq:baseband-signal}, we obtain the behaviour of the baseband deflection
\begin{align}
    \tilde{q}'_{\mathrm{b}}(s)
    &= \left\{
        \frac{1}{s}
        +
        \frac{1}{k} {G(s + i\omega)}
        \left[
            \delta\tilde{k}_{\mathrm{ts}}(s)
            - i\omega \, \delta\tilde{\gamma}_{\mathrm{ts}}(s)
        \right]
      \right\} \hat{q}_{\mathrm{s}} \nonumber \\
    &{}
      + \frac{1}{k} {G(s + i\omega)}
      \left[
        \delta\tilde{k}_{\mathrm{ts}}(s+2i\omega)
        - i\omega \, \delta\tilde{\gamma}_{\mathrm{ts}}(s+2i\omega)
      \right] \hat{q}^*_{\mathrm{s}} \nonumber \\
    &\phantom{=}
      + \frac{1}{k} {G(s + i\omega)} \, \left[
        \delta\tilde{F}_{\mathrm{ts}}(s+i\omega)
        + k \, \delta\tilde{a}_{\mathrm{exc}}(s)
      \right] \nonumber \\
    &\phantom{=}
      + i \left[\frac{G(s+i\omega)}{G(i\omega)} - 1\right]
        \delta\tilde{\phi}_{\mathrm{exc}}(s) \, \hat{q}_{\mathrm{s}}
    \label{eq:baseband-signal-explicit}
\end{align}

The resulting deflection can be viewed as a combined amplitude and phase
modulation
\begin{align}
    q(t)
        &=
          q_{\mathrm{s}}
          [1 + \delta m(t)]
          \cos(
              \omega t + \phi_{\mathrm{s}}
              + \delta\phi(t)
          ),
    \label{eq:deflection-perturbed}
\end{align}
where \(q_{\mathrm{s}} = |\hat{q}_{\mathrm{s}}|\) and \(\phi_{\mathrm{s}} = \arg
\hat{q}_{\mathrm{s}}\). For small modulations \(\delta m\) and
\(\delta \phi\) (narrowband approximation), the corresponding baseband
signal is
\begin{align}
    q'_{\mathrm{b}}(s)
        &= \left\{ \frac{1}{s}
        + \delta \tilde{m}(s)
        + i \delta \tilde{\phi}(s)
        \right\} \hat{q}_{\mathrm{s}}.
    \label{eq:baseband-mod}
\end{align}
We can furthermore introduce the complex modulation signal \(\delta
\tilde{\alpha} := \delta \tilde{m} + i \delta \tilde{\phi}\), which captures the
amplitude and phase modulation in its real and imaginary component,
respectively.
Comparing \cref{eq:baseband-mod} and \cref{eq:baseband-signal-explicit}, we obtain
\begin{align}
    \delta \tilde{\alpha} (s)
        &=
        \frac{1}{k} {G(s + i\omega)}
        \left[
            \delta\tilde{k}_{\mathrm{ts}}(s)
            - i\omega \, \delta\tilde{\gamma}_{\mathrm{ts}}(s)
        \right] \nonumber \\
        &{}
        +
        \frac{1}{k}
        \frac{\hat{q}^*_{\mathrm{s}}}{\hat{q}_{\mathrm{s}}}
        {G(s + i\omega)}
        \left[
            \delta\tilde{k}_{\mathrm{ts}}(s+2i\omega)
            + i\omega \, \delta\tilde{\gamma}_{\mathrm{ts}}(s+2i\omega)
        \right] \nonumber \\
        &{}\phantom{=}
        +
        \frac{1}{k a_0}
        \frac{G(s + i\omega)}{G(i \omega)} \left[
            \delta\tilde{F}_{\mathrm{ts}}(s+i\omega)
            + k \, \delta\tilde{a}_{\mathrm{exc}}(s)
        \right] \nonumber \\
        &\phantom{=}
        + i \left[\frac{G(s+i\omega)}{G(i\omega)} - 1\right]
          \delta\tilde{\phi}_{\mathrm{exc}}(s).
    \label{eq:complex-modulation}
\end{align}

We can further simplify \cref{eq:complex-modulation} if we assume an excitation
at the eigenfrequency \(\omega_0\). Then, the cantilever
response is phase-lagging at \(90^{\circ}\) and amplified by \(Q\), \(G(i\omega)
\approx G(i\omega_0) = -i Q\), and \(\hat{q}^*_{\mathrm{s}} /
\hat{q}_{\mathrm{s}} \approx -1\). The steady-state transfer function of the
cantilever \(G(s)\) can be approximated by a first-order system:
\begin{align}
    {G(s + i\omega)}
        \approx \frac{G(i\omega)}{1 + s / \omega_{\mathrm{c}}}
        \approx -\frac{i}{2} \frac{\omega_0}{\omega_{\mathrm{c}} + s}.
    \label{eq:transfer-function-approx}
\end{align}

In this limit, we obtain
\begin{align}
    \delta \tilde{\alpha} (s)
        &\approx
        \frac{1}{k a_0}
        \frac{\omega_{\mathrm{c}}}{\omega_{\mathrm{c}} + s}
        \left[
            \delta\tilde{F}_{\mathrm{ts}}(s+i\omega)
            + k \delta\tilde{a}_{\mathrm{exc}}(s)
        \right] \nonumber \\
        &\phantom{=}
        -\frac{\omega_0}{2 k} \frac{1}{\omega_{\mathrm{c}} + s}
        \left[
            i \delta\tilde{k}_{\mathrm{ts}}(s)
            + \omega \, \delta\tilde{\gamma}_{\mathrm{ts}}(s)
        \right] \nonumber \\
        &\phantom{=}
        +
        \frac{\omega_0}{2 k}
        \frac{1}{\omega_{\mathrm{c}} + s}
        \left[
            i \delta\tilde{k}_{\mathrm{ts}}(s+2i\omega)
            - \omega \, \delta\tilde{\gamma}_{\mathrm{ts}}(s+2i\omega)
        \right] \nonumber \\
        &\phantom{=}
        - i
        \frac{s}{\omega_{\mathrm{c}} + s}
          \delta\tilde{\phi}_{\mathrm{exc}}(s).
    \label{eq:complex-modulation-approx}
\end{align}

We can further split the frequency-shifted perturbations into components in- and
out-of-phase with the drive signal,
\begin{align}
    \delta\tilde{F}_{\mathrm{ts}} (s+i\omega)
        &= \delta\tilde{F}^{\omega,\mathrm{i}}_{\mathrm{ts}} (s)
         + i \delta\tilde{F}^{\omega,\mathrm{q}}_{\mathrm{ts}} (s), \\
    \delta\tilde{k}_{\mathrm{ts}} (s+2i\omega)
        &= \delta\tilde{k}^{2\omega,\mathrm{i}}_{\mathrm{ts}} (s)
         + i \delta\tilde{k}^{2\omega,\mathrm{q}}_{\mathrm{ts}} (s), \\
    \delta\tilde{\gamma}_{\mathrm{ts}} (s+2i\omega)
        &= \delta\tilde{\gamma}^{2\omega,\mathrm{i}}_{\mathrm{ts}} (s)
         + i \delta\tilde{\gamma}^{2\omega,\mathrm{q}}_{\mathrm{ts}} (s).
\end{align}

For the components of the amplitude and phase modulation, we obtain:
\begin{subequations}
\begin{align}
    \delta \tilde{m} (s)
        &\approx
        \frac{1}{\omega_{\mathrm{c}} + s}
        \bigg\{
            \frac{\omega_{\mathrm{c}}}{k a_0}
            \left[
                \delta\tilde{F}^{\omega,\mathrm{i}}_{\mathrm{ts}}(s)
                + k \, \delta\tilde{a}_{\mathrm{exc}}(s)
            \right]
            \nonumber \\
        &\hspace{2em}
            - \frac{\omega_0}{2 k}
            \left[
                \delta\tilde{k}^{2\omega,\mathrm{q}}_{\mathrm{ts}} (s)
                + \omega \, \delta\tilde{\gamma}_{\mathrm{ts}}(s)
                + \omega \, \delta\tilde{\gamma}^{2\omega,\mathrm{i}}_{\mathrm{ts}} (s)
            \right]
        \bigg\}
        \nonumber \\
    \label{eq:amplitude-modulation-approx} \\
    \delta \tilde{\phi} (s)
        &\approx
        \frac{1}{\omega_{\mathrm{c}} + s}
        \bigg\{
            \frac{\omega_{\mathrm{c}}}{k a_0} \delta\tilde{F}^{\omega,\mathrm{q}}_{\mathrm{ts}} (s) \nonumber \\
    &\hspace{3em}
            - \frac{\omega_0}{2 k}
            \left[
                \delta\tilde{k}_{\mathrm{ts}}(s)
                - \delta\tilde{k}^{2\omega,\mathrm{i}}_{\mathrm{ts}} (s)
                + \omega \, \delta\tilde{\gamma}^{2\omega,\mathrm{q}}_{\mathrm{ts}} (s)
            \right]
        \bigg\}
        \nonumber \\
    &\phantom{=}
        - \frac{s}{\omega_{\mathrm{c}} + s}
          \delta\tilde{\phi}_{\mathrm{exc}}(s).
    \label{eq:phase-modulation-approx}
\end{align}
\end{subequations}

The instantaneous frequency is defined as the phase derivative. Therefore, the
frequency modulation \(\delta \tilde{\omega}\) resulting from the phase
modulation in \cref{eq:phase-modulation-approx} can be derived as
\begin{align}
    \delta \tilde{\omega}(s)
        &= s \, \delta \tilde{\phi}(s) \nonumber \\
        &\approx
        \frac{s}{\omega_{\mathrm{c}} + s}
        \bigg\{
        \frac{\omega_{\mathrm{c}}}{k a_0}
          \delta\tilde{F}^{\mathrm{q}}_{\mathrm{ts}}(s+i\omega)
        - \delta\tilde{\omega}_{\mathrm{exc}}(s) \nonumber \\
        &\phantom{=}
        - \frac{\omega_0}{2 k}
        \left[
            \delta\tilde{k}_{\mathrm{ts}}(s)
            - \delta\tilde{k}^{2\omega,\mathrm{i}}_{\mathrm{ts}} (s)
            + \omega \, \delta\tilde{\gamma}^{2\omega,\mathrm{q}}_{\mathrm{ts}} (s)
        \right]
        \bigg\},
    \label{eq:frequency-modulation}
\end{align}
where \(\delta\tilde{\omega}_{\mathrm{exc}} = s
\delta\tilde{\phi}_{\mathrm{exc}}\) is the modulation of the drive frequency.

This allows for the following general observations:

\begin{itemize}
    \item The amplitude is modulated by changes of the excitation amplitude
    \(\delta\tilde{a}_{\mathrm{exc}}\), the damping coefficient
    \(\delta\tilde{\gamma}_{\mathrm{ts}}\), and its in-phase component at
    \(2\omega\). The amplitude can also be modulated by the in-phase component
    of a tip-sample force \(\delta\tilde{F}_{\mathrm{ts}}\) at \(\omega\) and
    the quadrature component of \(\delta\tilde{k}_{\mathrm{ts}}\) at
    \(2\omega\). These changes all pass through a lowpass filter at
    \(\omega_{\mathrm{c}}\).
    \item The phase and instantaneous frequency are modulated due to changes of
    the force gradient \(\delta\tilde{k}_{\mathrm{ts}}\), its in-phase component
    at \(2\omega\). It is also modulated by the quadrature component of
    \(\delta\tilde{F}_{\mathrm{ts}}\) at \(\omega\) and the quadrature component
    of \(\delta\tilde{\gamma}_{\mathrm{ts}}\) at \(2\omega\). For the detected
    phase, these changes all pass through a lowpass filter, whereas the
    excitation phase enters via a highpass filter at \(\omega_{\mathrm{c}}\).
    For the instantaneous frequency, there is a highpass filter at
    \(\omega_{\mathrm{c}}\) common to all inputs.
\end{itemize}

A widespread misconception, dating back to the time when the FM-AFM technique
was first introduced by \citet{Albrecht:1991bu}, is that phase and frequency
changes propagate instantaneously. This statement is obviously true only to a
limited extent. Especially in ambient or liquid environments, quality factors
are typically well below \num{1000}, and the cutoff frequency can be on the
order of hundreds of Hertz. Phase changes due to interactions with the surface
are then detected easily within the bandwidth \(\omega_{\mathrm{c}}\), whereas
changes of the excitation phase can cause strong transients due to the highpass
characteristic. On the other hand, in high vacuum quality factors of \(\sim
100000\) can be achieved, resulting in cutoff frequencies on the order of
\SI{1}{Hz}. When slow processes are observed, or when feedback loops are
intentionally kept at low speeds, the exact expression should be used also under
vacuum conditions. Only for quality factors \(Q\) approaching infinity, the
cutoff frequency is negligible, and phase changes of the excitation affect the
cantilever oscillation nearly immediately.

\section{Transfer functions for FM-AFM}

Next, we derive transfer functions relevant for FM-AFM operation.
\Cref{eq:complex-modulation} contains the behavior of the amplitude and phase
modulations as a function of the driving amplitude and phase of the cantilever.
These are the transfer functions of amplitude and phase, \(G_{\mathrm{m}}(s)\)
and \(G_{\phi}(s)\), respectively,
\begin{align}
    G_{\mathrm{m}}(s)
        &:= \frac{\delta \tilde{m} (s)}{\delta\tilde{a}_{\mathrm{exc}}(s)}
        = \frac{1}{a_0} \frac{G(s+i\omega)}{G(i\omega)}
        \approx \frac{1}{a_0} \frac{\omega_{\mathrm{c}}}{\omega_{\mathrm{c}} + s}
        \label{eq:transfer-function-am}\\
    G_{\phi}(s)
        &:= \frac{\delta \tilde{\phi} (s)}{\delta\tilde{\omega}_{\mathrm{exc}} (s)}
        = \frac{1}{s} \left[\frac{G(s+i\omega)}{G(i\omega)} - 1\right]
        \approx
        - \frac{1}{\omega_{\mathrm{c}} + s}.
        \label{eq:transfer-function-fm}
\end{align}
The approximations following from
\cref{eq:amplitude-modulation-approx,eq:phase-modulation-approx} are valid for
an excitation close to \(\omega_0\) and assume that \(\omega_{\mathrm{c}} \ll
\omega_0\), such that the cantilever transfer function can be considered a
first-order system.

As indicated by \cref{eq:transfer-function-am}, amplitude changes propagate via
a low-pass filter at \(\omega_{\mathrm{c}}\). Changes of the drive frequency
propagate via \cref{eq:transfer-function-fm} to the detected cantilever phase,
which has the characteristics of an integrator with a cutoff frequency
\(\omega_{\mathrm{c}}\). As discussed below, this behaviour has important
implications for the design of a phase-locked loop.

In the following, we use \cref{eq:transfer-function-am,eq:transfer-function-fm}
to find the closed-loop behaviour of the feedback loops governing FM-AFM
performance and stability.

\subsection{Phase-locked loop}

A PLL is used to track the resonance frequency and to excite the cantilever
accordingly\cite{Durig:1992cn,Durig:1997en,Loppacher:1998gr}. The output signal
of the PLL is the frequency shift \(\Delta \omega\), which is used as the input
signal to the topography feedback loop. Following \cref{eq:transfer-function-fm}
every change of the instantaneous resonance frequency has an effect on the
detected phase. Therefore, by adjusting the excitation frequency to maintain a
constant phase, the cantilever can be driven at resonance.

A PLL is composed of several building blocks: a tunable reference oscillator, a
phase detector to measure the phase relation of the input oscillation to the
reference, and a loop filter responsible for closed-loop control.

The phase detector is commonly built by down-conversion of the input signal by
the reference oscillator, followed by low-pass filter to remove high-frequency
mixing products. To achieve higher bandwidths up to the cantilever
eigenfrequency, alternative phase detectors have been proposed
recently\cite{Mitani:2009ca,Schlecker:2014kf,Miyata:2018gt}.

For small phase shifts, the phase detector can be approximated by the transfer
function of the filter \(F_\phi(s)\). It should be mentioned that the phase
detector is no longer linear for the large phase shifts which appear initially
when acquiring the phase lock\cite{Best:2003tz}. Next, we therefore examine the
performance in the phase-locked state only.

\begin{figure}
    \includegraphics[width=\linewidth]{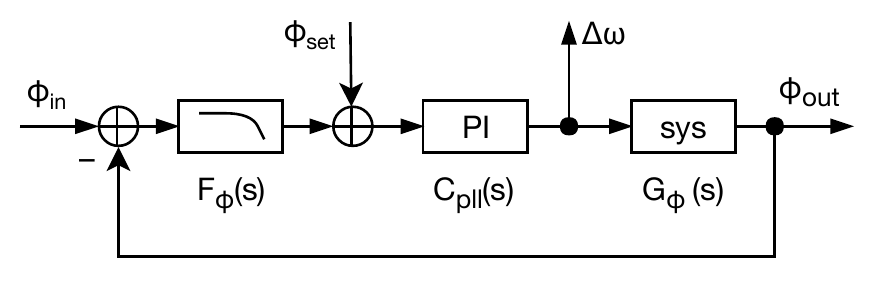}
    \caption{Block diagram of the phase-locked loop.}
    \label{fig:pll-blockdiagram}
\end{figure}
The block diagram in \cref{fig:pll-blockdiagram} illustrates the complete phase
feedback loop. To maintain the phase at resonance (setpoint
\(\phi_{\mathrm{set}} = \SI{-90}{\degree}\)), the drive frequency is adjusted by
a proportional--integral (PI) controller, \(C_{\mathrm{pll}}\). Phase changes of
the cantilever, which describe the deviations from resonance, enter the system
via the input phase \(\phi_{\mathrm{in}}\), whereas the resulting controlled
phase is given by \(\phi_{\mathrm{out}}\). The closed-loop transfer functions of
the PLL are:
\begin{align}
    \tilde{\phi}_{\mathrm{out}}
        &= G_{\phi} C_{\mathrm{pll}}
        \left[
            \phi_{\mathrm{set}}
            + F_{\phi} \left( \tilde{\phi}_{\mathrm{in}} - \tilde{\phi}_{\mathrm{out}} \right)
        \right ]
        \nonumber \\
    \Rightarrow \,
        \frac{\tilde{\phi}_{\mathrm{out}}}{\tilde{\phi}_{\mathrm{in}}}
        &=  \frac{ G_{\phi} C_{\mathrm{pll}} F_{\phi} }
            {1 + G_{\phi} C_{\mathrm{pll}} F_{\phi}},
    \;
        \frac{\tilde{\phi}_{\mathrm{out}}}{\tilde{\phi}_{\mathrm{set}}}
        =  \frac{ G_{\phi} C_{\mathrm{pll}} }
            {1 + G_{\phi} C_{\mathrm{pll}} F_{\phi}}.
    \label{eq:closed-loop-fm}
\end{align}

Note that the transfer functions for \(\tilde{\phi}_{\mathrm{in}}\) and
\(\tilde{\phi}_{\mathrm{set}}\) are slightly different, because changes of the
setpoint are not affected by the phase detection filter \(F_\phi\). Care should
be taken when testing the response of the feedback loop experimentally by
changing the setpoint, since there might be deviations from the closed-loop
response rejecting disturbances \(\phi_{\mathrm{in}}\).

The dynamic performance of the feedback loop is determined by the poles of the
PLL transfer function, which are the zeros of \(1 + G_{\mathrm{ol}}\), where
\(G_{\mathrm{ol}} = G_{\phi} C_{\mathrm{pll}} F_{\phi}\) is the open-loop
transfer function.

For \(C_{\mathrm{pll}}(s)\), we can write
\begin{align}
    C_{\mathrm{pll}}(s)
        &= K_{\mathrm{p}} + \frac{K_{\mathrm{i}}}{s}
        = K_{\mathrm{p}} \frac{ \omega_{\mathrm{pi}} + s }{ s },
    \label{eq:transfer-function-pi}
\end{align}
where \(K_{\mathrm{p}}\) and \(K_{\mathrm{i}}\) are the proportional and
integral gains of the feedback loop, and \(\omega_{\mathrm{pi}} = K_{\mathrm{i}}
/ K_{\mathrm{p}}\). Therefore, in the open-loop transfer function
\(G_{\mathrm{ol}}\) the pole at \(\omega_{\mathrm{c}}\) due to
\cref{eq:transfer-function-fm} can be cancelled by the zero at
\(\omega_{\mathrm{pi}}\) due to \cref{eq:transfer-function-pi} by choosing
appropriate feedback parameters.

Furthermore, the poles and zeros of the filter \(F_{\phi}\) can be chosen such
that they are well above the desired closed-loop bandwidth. In this limit, the
closed-loop transfer function of the PLL is the response of a first-order
lowpass filter
\begin{align}
    \frac{\tilde{\phi}_{\mathrm{out}}}{\tilde{\phi}_{\mathrm{in}}}
    &\approx
    \frac{ \omega_{\mathrm{cl}} }{ s + \omega_{\mathrm{cl}} },
\end{align}
where the closed-loop bandwidth is \(\omega_{\mathrm{cl}} = -K_{\mathrm{p}}\).
Note that the proportional gain \(K_{\mathrm{p}}\) must be negative to obtain a
stable closed-loop system under negative feedback. This is because following
\cref{eq:transfer-function-fm} an increased drive frequency results in a reduced
phase response.

PLL transfer functions in literature\cite{Loppacher:1998gr,Kim:2004ev} are
commonly derived under the assumption of negligible damping, i.e.,
\(\omega_{\mathrm{c}} \rightarrow 0\). The detected phase in
\cref{eq:transfer-function-fm} is then merely an integral of frequency, and a
first-order PLL defined by its proportional gain alone would suffice for good
performance. Although in ultra-high vacuum this assumption holds very
well\cite{Loppacher:1998gr}, with lower quality factors as typical in ambient or
liquid environments there can be significant deviations which can cause
experimental artifacts\cite{Kaggwa:2008gl,Fukuma:2011dg}.

Note that most textbooks\cite{Best:2003tz,Gardner:2005tj,Egan:2007wb} deal with
PLLs in the context of signals and communications systems, where the detected
phase is indeed the integrated change of the reference frequency. In the context
of AFM this situation is found in combination with a self-excitation
setup\cite{Albrecht:1991bu}, in which the detected deflection of the cantilever
is delayed or phase-shifted, amplified, and applied as an excitation signal. For
this configuration, a proportional-type controller is sufficient, because the
measured frequency shift is detected only and not fed back via the excitation
signal.

For PLL-based excitation, in contrast, the cantilever transfer function enters
the excitation loop and must be considered. The main advantage of PLL-based
excitation is the clean driving signal derived from the narrow-band phase
detector, which enables working with low-Q cantilevers\cite{Durig:1997en} and
provides better noise performance\cite{Kim:2004ev}. It is unfortunate that
several publications focusing on FM-AFM instrumentation and simulation initially
neglected the integral feedback required in the general
case\cite{Loppacher:1998gr,PoleselMaris:2005he,Nony:2006bo}. Related to that,
the phase response of the cantilever must necessarily be considered within the
phase feedback loop. Attempts to incorporate the phase response otherwise, e.g.,
from an outer feedback loop\cite{Durig:1997en}, or treating the combined system
as interlaced control loops\cite{Lubbe:2016ec} have resulted in contradictory
transfer functions.

To estimate the error with a proportional-only controller (\(C_{\mathrm{pll}} =
K_{\mathrm{p}}\)), we derive the closed-loop transfer function, omitting the
phase detector, as
\begin{align}
    \frac{\tilde{\phi}_{\mathrm{out}}}{\tilde{\phi}_{\mathrm{in}}}
    &\approx
    \frac{ 1 }{ 1 - \omega_{\mathrm{c}} / K_{\mathrm{p}} }
    \frac{ \omega_{\mathrm{c}} - K_{\mathrm{p}} }{ s + \omega_{\mathrm{c}} - K_{\mathrm{p}} },
    \label{eq:pll-closed-loop-onlyp}
\end{align}
corresponding to a lowpass filter with the cutoff frequency
\(\omega_{\mathrm{c}} - K_{\mathrm{p}}\) and gain reduced by a factor of \((1 -
\omega_{\mathrm{c}} / K_{\mathrm{p}})\). If \(\omega_{\mathrm{c}}\) is not
negligible compared to \(K_{\mathrm{p}}\), a proportional-only controller can
lead to an apparent reduction of the frequency shift \(\Delta \omega\) and
thereby cause inaccurate FM-AFM measurements.

\begin{figure}
    \includegraphics[width=\linewidth]{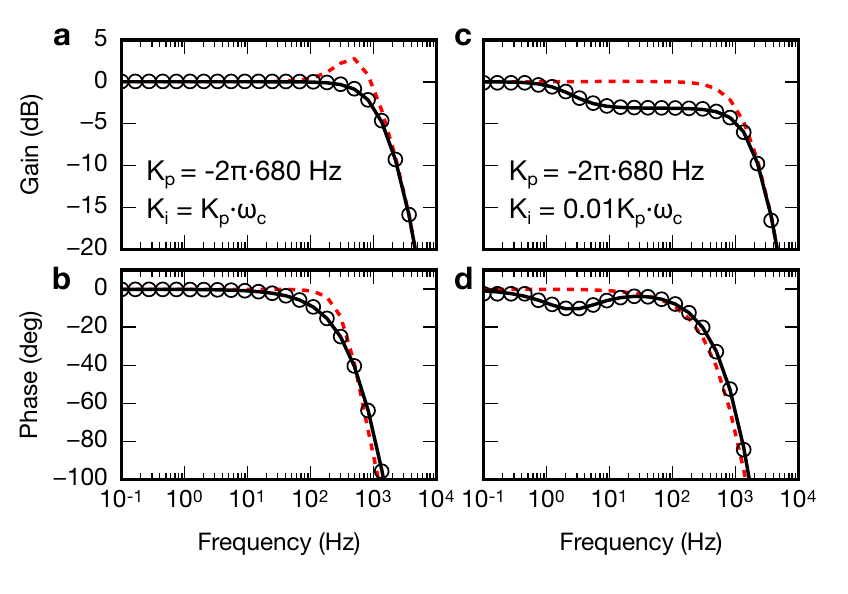}
    \caption{
        Influence of the integral gain on the closed-loop behaviour of the PLL.
        The black lines indicate the response as calculated from
        \cref{eq:closed-loop-fm}. The circles show a full numerical simulation
        based on the cantilever dynamics and
        detection system. The response for an ideal oscillator, i.e.,
        \(\omega_{\mathrm{c}}=0\), is indicated by a dashed red line. A
        cantilever with \(\omega_0=2\pi\,\SI{300}{kHz}\) and \(Q=500\) is
        assumed, and the phase detector is modeled as a four-pole lowpass with a
        \SI{-3}{dB} bandwidth of \SI{4}{kHz}. (a)~Gain and (b)~phase response for
        a PLL tuned to a bandwidth of \SI{1}{kHz}. (c)~Gain and (d)~phase
        response with integral gain reduced by a factor of \(100\).
    }
    \label{fig:pll-simulation}
\end{figure}

\Cref{fig:pll-simulation} shows the influence of the integral gain on the
closed-loop behaviour of the PLL for a real (\(\omega_0=2\pi\,\SI{300}{kHz}\),
\(Q=500\)) and ideal oscillator (\(\omega_{\mathrm{c}}=0\)).
Also shown are the results of a simulation of the cantilever dynamics and the
detection system. For the simulation, the time is discretized to \(\Delta
t=(\SI{1}{MHz})^{-1}\), and the equation of motion,
\cref{eq:eq-of-motion-approx}, is solved by velocity Verlet integration with 10
intermediate steps. At each time step \(\Delta t\), the phase is detected by a
lock-in amplifier and maintained at \SI{-90}{\degree} by a PI controller. To
obtain each point in the simulation, the resonance frequency is modulated at a
single frequency \(\omega_{\mathrm{m}}\), and the resulting frequency shift
signal is recorded during at least ten cycles. Gain and phase are calculated
from the Fourier coefficient of the frequency shift signal at
\(\omega_{\mathrm{m}}\).

The simulation shows that the closed-loop behaviour follows
\cref{eq:closed-loop-fm} very well. As described above, the closed-loop
bandwidth scales with \(K_{\mathrm{p}}\). A hundredfold reduction of the
integral gain results in a reduction of the amplitude at higher frequencies by
nearly \(\SI{30}{\percent}\) due to dominant proportional-type action. In an
actual experiment, a controller tuned similar to \cref{fig:pll-simulation}(c)
and (d) can be detrimental. To the operator, the feedback loop appears to work
properly, because there is no static error and the frequency shift \(\Delta
\omega\) appears to follow up to the desired bandwidth. It is easiest to detect
such a situation experimentally from a step response, for example by retracting
the tip from the surface. With too little integral action, \(\Delta \omega\)
would slowly creep to the final value.

\subsection{Amplitude controller}
Detected amplitude changes \(\delta \tilde{A}\) are determined by the modulation
\(\delta \tilde{m}\) and the steady-state cantilever amplitude
\(A = |\hat{q}_{\mathrm{s}}|\),
\begin{align}
    G_{\mathrm{A}}(s)
        &= \frac{\delta\tilde{m}(s) A}{\delta\tilde{a}_{\mathrm{exc}}(s)}
        = G_{\mathrm{m}}(s) A
        \approx Q \frac{ \omega_{\mathrm{c}} }{ \omega_{\mathrm{c}} + s },
\end{align}
where the approximation holds for excitation at resonance.
Similar to the derivation of the PLL transfer function, the closed-loop transfer
function of the amplitude controller is obtained as
\begin{align}
    \frac{ \tilde{A}_{\mathrm{out}}}{ \tilde{A}_{\mathrm{in}}}
        &=  \frac{ G_{\mathrm{A}} C_{\mathrm{A}} F_{\mathrm{A}} }
        {1 + G_{\mathrm{A}} C_{\mathrm{A}} F_{\mathrm{A}}},
    \label{eq:closed-loop-am}
\end{align}
where \(F_{\mathrm{A}}(s)\) is the transfer function of the amplitude detection
filter and a PI controller \(C_{\mathrm{A}}(s)\) is used to maintain a constant
amplitude. Choosing \(\omega_{\mathrm{pi}}\) again to compensate the dominating
pole at \(\omega_{\mathrm{c}}\), and neglecting the filter \(F_{\mathrm{A}}\),
we obtain
\begin{align}
    \frac{ \tilde{A}_{\mathrm{out}} }{ \tilde{A}_{\mathrm{in}} }
    &\approx
    \frac{ \omega_{\mathrm{cl}} }{ s + \omega_{\mathrm{cl}} }
    \quad \text{with} \quad
    \omega_{\mathrm{cl}} = K_{\mathrm{p}} Q \omega_{\mathrm{c}} = K_{\mathrm{p}} \omega_0 / 2.
\end{align}

\section{Discussion}

The relations derived in \cref{sec:transients,sec:baseband} together with the
steady-state solution from \cref{sec:steadystate} describe the full input/output
relationship of interaction and driving forces acting on the deflection of the
AFM tip.

The solution for perturbations of the deflection
\(\delta\tilde{q}_{\mathrm{int}}\), \cref{eq:delta-q-int-explicit}, shows that
changes of the interaction \(\delta \tilde{F}_{\mathrm{ts}}\) enter at low
frequencies without change, because the transfer function \(G(s)\) at low
frequencies is unity. Particularly, there is no additional lowpass filtering,
even if the common designation \emph{cantilever bandwidth} for
\(\omega_{\mathrm{c}}\) would suggest so. Modulation at the drive frequency
\(\omega\) in \(\delta\tilde{k}_{\mathrm{ts}}\) or
\(\delta\tilde{\gamma}_{\mathrm{ts}}\) can also be detected immediately in the
deflection signal, that is, without prior demodulation.

The complex baseband signal, \cref{eq:complex-modulation}, describes the
cantilever response at the driving frequency \(\omega\). In our derivation, we
consider weak perturbations around the steady-state solution. Besides, there are
no assumptions about the quality factor of the cantilever or the deviation of
the drive frequency from resonance. Hence, this solution is applicable even in
environments of high damping.

\Cref{eq:amplitude-modulation-approx,eq:phase-modulation-approx,eq:frequency-modulation}
describe the amplitude, phase, and frequency response in an environment of
moderate damping when driving close to resonance. In particular, these equations
reflect the well-known fact that the phase and frequency modulation of the
cantilever is primarily due to a modulation of the force gradient, whereas the
amplitude is modulated due to dissipation. Moreover, amplitude and phase are
modulated by the in-phase and quadrature components of the driving force,
respectively.

In addition, in-phase and quadrature components of \(\delta
\tilde{k}_{\mathrm{ts}}\) or \(\delta \tilde{\gamma}_{\mathrm{ts}}\) at
\(2\omega\) also modulate the amplitude and phase of the deflection at
\(\omega\). Such heterodyne modulation and detection offers an interesting route
to probe tip--sample interactions at high frequencies. The effect has been
utilized, for example, to implement a force gradient sensitive detection method
for Kelvin probe force microscopy using the dissipation
channel\cite{Miyahara:2017vg}. Because the dissipation channel, contained in
\(\delta m\), is orthogonal to the frequency modulation \(\delta \omega\), a
larger Kelvin detection bandwidth can be achieved compared to the traditional
approach of modulating \(\delta k_{\mathrm{ts}}\) at low frequencies. A downside
to this approach is that amplitude changes due to increased \emph{real}
dissipation or changes of the driving behaviour may be mistaken for changes of
the measured surface potential.

Higher harmonic terms in \cref{eq:delta-q-int-explicit} appear due to the highly
non-linear tip--sample force even if the tip oscillation can be considered
harmonic. In practice, they are difficult to detect off-resonance at frequencies
\(n\omega\). However, a modulation of the tip--sample force at a frequency
\(\omega_{\mathrm{m}}\) also modulates the harmonics \(\delta \tilde{f}_{\pm
n}\) at \(\omega_{\mathrm{m}}\). Frequency mixing with harmonics of the carrier
oscillation can therefore be used to excite the cantilever at frequencies
\(n\omega + \omega_{\mathrm{m}}\). Through appropriate choice of the order \(n\)
and modulation frequency \(\omega_{\mathrm{m}}\), it is possible to amplify the
resulting signal by an eigenmode of the cantilever. This approach has been
termed \emph{harmonic mixing} and was exploited recently to tune the imaging
resolution in Kelvin probe force microscopy\cite{Garrett:2018ck}. In case of the
fundamental harmonic, \(n=1\), harmonic mixing is equivalent to
heterodyne\cite{Sugawara:2012bh,Garrett:2016dn} and sideband
(de)modulation\cite{Zerweck:2005kc,Wagner:2015ci} techniques.

The transfer functions derived from the baseband response of the cantilever are
indispensable in the design of feedback loops. Empirical tuning by trial and
error should be avoided, even more so when feedback loops depend on each other.
In case of PLL-based excitation, it is crucial to include the influence of the
cantilever for low quality factors.

The separation of cantilever response into a steady-state and transient
solution calls for an obvious application in the optimization of AFM simulators.
Such
simulators\cite{Couturier:2001ka,PoleselMaris:2005he,Nony:2006bo,Trevethan:2007ct,Kiracofe:2012ft,Tracey:2015eh}
are used to understand the influence of feedback loops and the operator on the
measurement, can explain imaging artifacts\cite{Nony:2009bi,Nony:2009iv} and
correlate experiment and theory\cite{Castanie:2013gc}. Moreover, they have a
potential use in the training and education of new AFM
users\cite{Kiracofe:2012ft,Guzman:2015dl}.

There is a tradeoff between accuracy of the simulation and the computational
power required. The prevalent
approach\cite{PoleselMaris:2005he,Nony:2006bo,Kiracofe:2012ft,Tracey:2015eh} is
the direct numerical solution of the equation of motion, \cref{eq:eq-of-motion}.
While being the most general and accurate, however, the direct solution is
computationally very demanding, because for numerical stability the time step
must be chosen significantly below the oscillation period. On the other hand,
the static solution alone, given by
\cref{eq:afm-Fts,eq:afm-kts,eq:kts-k-approx,eq:afm-gamma}, already allows one to
calculate AFM images under ideal conditions, i.e., neglecting the effects of
feedback loops and possible modifications of the AFM tip and surface.
The dynamics can be considered on top of the static solution by a numerical
solution of the baseband equation of motion,
\cref{eq:baseband-equation-of-motion}, which accurately models the cantilever
dynamics relevant for AM- and FM-AFM. Since high-frequency dynamics are
neglected in this equation, the time steps for the integration can be chosen
much larger compared to direct integration of \cref{eq:eq-of-motion}.
Additionally, for numerical investigations, it is often possible to calculate
\(\langle k_{\mathrm{ts}} \rangle\) and \(\langle \gamma_{\mathrm{ts}} \rangle\)
in advance. For this reason, an AFM simulator built around
\cref{eq:baseband-equation-of-motion} can be implemented very efficiently.

\section{Summary}

This work provides compact solutions of the steady-state and transient behavior
of an AFM tip under the influence of external forces. Whereas the steady-state
behavior has been studied extensively before for AM-AFM and FM-AFM individually,
a unifying derivation of the underlying solution based on the harmonic
approximation was only given recently\cite{Songen:2017kc}.

We have derived the transient behavior of the deflection signal from a
perturbation of this steady-state solution. In essence, it was shown that
transients are linked to the time evolution of Fourier coefficients of the
tip--sample force and their modulation by the cantilever response function.

We have also described the complex baseband behavior and provided expressions of
the resulting amplitude, phase, and frequency modulation. These were used to
derive the transfer functions of the PLL and amplitude controller used for
FM-AFM.

Our results show that a holistic treatment of the cantilever movement enables a
deeper understanding of its behavior and reveals the many interconnections and
similarities of dynamic AFM techniques used today and in future.

\begin{acknowledgments}
    The author greatly appreciates the fruitful and constructive discussions
    with Prof.\ Andreas Stemmer throughout the preparation of this manuscript.
\end{acknowledgments}

\bibliography{references}

\end{document}